\documentclass[%
 reprint,
 showpacs,preprintnumbers,
 amsmath,amssymb,
 aps,
 pre,
floatfix,
]{revtex4-1}

\usepackage{graphicx, verbatim, multirow, enumitem}
\usepackage[usestackEOL]{stackengine}
\usepackage{dcolumn}% Align table columns on decimal point
\usepackage{bm}% bold math
\usepackage[caption=false]{subfig}
\newcommand{\subfigimg}[3][,]{%
  \setbox1=\hbox{\includegraphics[#1]{#3}}% Store image in box
  \leavevmode\rlap{\usebox1}% Print image
  \rlap{\hspace*{10pt}\raisebox{\dimexpr\ht1-2\baselineskip}{#2}}% Print label
  \phantom{\usebox1}% Insert appropriate spcing
}
\begin{document}

\preprint{APS/123-QED}

\title{Topology-dependent rationality and quantal response equilibria\\ in structured populations}% Force line breaks with \\

\author{Sabin Roman}
 \email{sr10g13@soton.ac.uk}%Lines break automatically or can be forced with \\
\author{Markus Brede}%
\affiliation{%
Agents, Interaction and Complexity Group,\\
 School of Electronics and Computer Science,\\
 University of Southampton, UK %\textbackslash\textbackslash
}%

\collaboration{Institute for Complex Systems Simulation}\noaffiliation

\date{\today}% It is always \today, today,

\begin{abstract}

Given that the assumption of perfect rationality is rarely met in the real world, we explore a graded notion of rationality in socioecological systems of networked actors. We parametrize an actors' rationality via their place in a social network and quantify system rationality via the average Jensen-Shannon divergence between the games Nash and logit quantal response equilibria. Previous work has argued that scale-free topologies maximize a system's overall rationality in this setup. Here we show that while, for certain games, it is true that increasing degree heterogeneity of complex networks enhances rationality, rationality-optimal configurations are not scale-free. For the Prisoner's Dilemma and Stag Hunt games, we provide analytic arguments complemented by numerical optimization experiments to demonstrate that core-periphery networks composed of a few dominant hub nodes surrounded by a periphery of very low degree nodes give strikingly smaller overall deviations from rationality than scale-free networks. Similarly, for the Battle of the Sexes and the Matching Pennies games, we find that the optimal network structure is also a core-periphery graph but with a smaller difference in the average degrees of the core and the periphery. These results provide insight on the interplay between the topological structure of socioecological systems and their collective cognitive behavior, with potential applications to understanding wealth inequality and the structural features of the network of global corporate control.

\end{abstract}

\pacs{02.50.Le, 87.23.Ge, 87.23.Kg, 89.75.Fb}% PACS, the Physics and Astronomy
\maketitle

%\tableofcontents

\section{Introduction}

Game theory has been a popular tool for modelling decision making scenarios, with applications that vary from understanding economic market forces \citep{Cardell1997}, to evolutionary biology \citep{Nowak2004}, to political science and considerations of warfare \citep{DeMesquita2006}. The identification of equilibrium states is a primary concern in the field, a key notion being that of Nash equilibria \cite{Nash1950}, which represent fully rational outcomes in strategy selection. Nevertheless, actors in the real world deviate from rational behaviour \citep{Nell2013}.  

As a consequence, extended notions of equilibrium are needed to deal with the discrepancies between real and perfectly rational behaviour, and to precisely quantify such bounded rationality. One alternative equilibrium concept is the notion of quantal response equilibrium (QRE), which is given by statistical reaction functions that satisfy certain monotony properties with respect to the payoffs of possible strategies \citep[p. 10]{McKelvey1995}. One possible class of parametric class of functions, which relate to the study of individual choice behaviour \citep{Luce2005}, define the logit quantal response equilibrium (LQRE). In LQRE a rationality parameter is used to interpolate between Nash equilibria and a uniform distribution over all the possible strategies \cite{McKelvey1995}. An early use of LQRE was to account for experimentally observed behaviour in choosing strategies when playing the Matching Pennies game \cite{Ochs1995, McKelvey1995}. Since then, the LQRE has found applications in the traditional areas of game theory, including auction theory \citep{Goeree2002}, participation games, e.g., voting \cite{Enriqueta2004, McKelvey2006}, and theories of social learning \citep{Rogers2009, Choi2012}. Here, we will make use of the LQRE in conjuncture with network theory.

Games on networks have had a rich history, with early research started by considering propagation of strategic behaviour in spatial settings, e.g, on lattices by considering the Prisoner's Dilemma \citep{Nowak1992} or, recently, the more general case of potential games \citep{Szabo2016}. In recent times, evolutionary games on networks have been a topic of growing interest, see, e.g., \citep{Szabo2007,Perc2013} for reviews. 

Topology can have a significant effect on equilibria of the system, e.g., scale-free networks have been found to promote cooperation in the Prisoner's Dilemma, provided that pairwise imitation is used with the probability of imitation proportional to the difference in total payoffs  \cite{Santos2005}. Other research has investigated the coevolution of internal states and topological features of networks where players are able to modify the structure by rewiring links, leading to a highly cooperative stationary state \citep{Zimmermann2004}, or, if cooperation is to be maintained as a viable strategy in the Prisoner's Dilemma then the network tends to acquire an exponential degree distribution \citep{Szolnoki2008}.

Also investigating the topological evolution of networks, the authors of \cite{Kasthurirathna2015} account for the fact that individuals are known to be influenced and biased by their social connections and thus consider LQREs for games played between agents connected via a social network. As more highly connected players are assumed to be in a better position to gather information, the rationality of players is modelled as an increasing function of node degree. The network structure is then optimised to minimise the overall deviation from perfect rationality (i.e., the Nash equilibrium), and the general result arrived at by \cite{Kasthurirathna2015} is that scale-free networks realise an optimum.

Furthermore, the scale-free topology seems to emerge independently of the games considered by \cite{Kasthurirathna2015}. Since then, the framework and results of \cite{Kasthurirathna2015} have been used in problems relating to internet routing \citep{Kasthurirathna2016a} and optimising influence in social networks \citep{Kasthurirathna2016b}. Here, we argue that the optimal topology to maximise rational behaviour among all players depends on the games under consideration, and in no case does it correspond to a scale-free network.

Below we show that for the Prisoner's Dilemma and the Stag Hunt games the optimum topology is a core-periphery network, where the core consists of a low number of high degree hubs and the periphery is made up of many low degree nodes. For the Battle of the Sexes and Matching Pennies games, the optimum topology is also of core-periphery type, but there are differences in the number of hubs, as well as in the average degree of the nodes in the core and in the periphery. In all cases we provide analytic treatment and perform computational experiments to support our claims. In Section \ref{sec:mm} we outline our methodology, which is closely aligned to that of \cite{Kasthurirathna2015}. In Section \ref{sec:rs} we present results, with Subsection \ref{subsec:pdsh} containing the treatment of the Prisoner's Dilemma and Stag Hunt games and  Subsection \ref{subsec:bsmp} dealing with the Battles of the Sexes and Matching pennies. Section \ref{sec:cc} concludes the paper.

\section{Methodology}
\label{sec:mm}

\subsection{Logit Quantal Response Equilibrium}

\bgroup
\def\arraystretch{1.5}% 
\begin{table}[ht]
\centering
\begin{tabular}{|c|c|c|}
\hline
Game                & \multicolumn{2}{c|}{P2}                             \\ \hline
\multirow{2}{*}{P1} & $u^{1}_{11}, u^{2}_{11}$ & $u^{1}_{12}, u^{2}_{12}$ \\ \cline{2-3} 
                    & $u^{1}_{21}, u^{2}_{21}$ & $u^{1}_{22}, u^{2}_{22}$ \\ \hline
\end{tabular}
\caption{Payoffs in the general form of two-person game.}
\label{tab:general}
\end{table}
\egroup

In this section we briefly revisit the framework of \citep{Kasthurirathna2015}. Parameter choices and the nature of the equilibria we analyse in the rest of the paper are closely aligned with \citep{Kasthurirathna2015} to facilitate comparison. Given a network, every node $i$ is assigned a rationality $\lambda_i = f(d_i)$ which is a function of the nodes' degree $d_i$. The rationality function is an increasing function of the degree, which implies that higher degree nodes have higher rationality. We consider the cases in which the function $f$ is: linear, where $f(d) = rd$ and $r = 0.2$, convex, where $f(d) = rd^{2}$ and $r = 0.002$ or concave, where $f(d) = r\sqrt{d}$ and $r = 0.5$.

A two-person game is played on every edge of the network, where the nodes represent the players and links denote the games. The payoffs for any two person game can be written as in Table \ref{tab:general}. We will be considering games such as the Prisoner's Dilemma and we are interested in determining the Logit Quantal Response Equilibrium (LQRE) of the games, rather than their Nash equilibrium. Each edge has two players, labeled 1 and 2, with rationality $\lambda_{1}$ and $\lambda_{2}$, respectively. We consider mixed strategies, where each player has two available pure strategies, labeled cooperation and defection. Player 1 cooperates with probability $p_{c}^{1}$ and defects with probability $1-p_{c}^{1}$, and similarly for player 2. Following \cite{McKelvey1995} the LQRE is determined by solving the system of equations

\begin{equation}
\begin{aligned}
p_{c}^{1} &= \cfrac{e^{\lambda_{1}(p_{c}^{2}u^{1}_{11} + (1-p_{c}^{2})u^{1}_{12})}}{e^{\lambda_{1}(p_{c}^{2}u^{1}_{11} + (1-p_{c}^{2})u^{1}_{12})} + e^{\lambda_{1}(p_{c}^{2}u^{1}_{21} + (1-p_{c}^{2})u^{1}_{22})}}\\
p_{c}^{2} &= \cfrac{e^{\lambda_{2}(p_{c}^{1}u^{2}_{11} + (1-p_{c}^{1})u^{2}_{21})}}{e^{\lambda_{2}(p_{c}^{1}u^{2}_{11} + (1-p_{c}^{1})u^{1}_{21})} + e^{\lambda_{2}(p_{c}^{2}u^{2}_{12} + (1-p_{c}^{2})u^{2}_{22})}}
\end{aligned}
\label{eq:qre}
\end{equation}

Alternatively, Eq.'s \eqref{eq:qre} can be written as:

\begin{equation}
\begin{aligned}
p^{1}_{c}(1+a_{1}b_{1}^{p^{2}_{c}}) &= 1\\
p^{2}_{c}(1+a_{2}b_{2}^{p^{1}_{c}}) &= 1
\end{aligned}
\label{eq:qrealt}
\end{equation}

where

\begin{equation}
\begin{aligned}
a_{1} &= \exp(\lambda_{1}(u^{1}_{22}-u^{1}_{12}))\\
b_{1} &= \exp(\lambda_{1}(u^{1}_{12}+u^{1}_{21}-u^{1}_{11}-u^{1}_{22}))\\
a_{2} &= \exp(\lambda_{2}(u^{2}_{22}-u^{2}_{21}))\\
b_{2} &= \exp(\lambda_{2}(u^{2}_{12}+u^{2}_{21}-u^{2}_{11}-u^{2}_{22}))
\end{aligned}
\label{eq:qreaux}
\end{equation}

Eq. \eqref{eq:qrealt} will be helpful when solving for different games. In case of the Prisoner's Dilemma, if $\lambda_{1}, \lambda_{2} \rightarrow \infty$, then $p^{1}_{c}, p^{2}_{c} \rightarrow 0$, which corresponds to the Nash equilibrium, i.e., a rational outcome. For an edge $k$ we write the probability distribution associated to player 1 as $P^{1}_{k} = (p_{c}^{1}, 1-p_{c}^{1})$, and likewise for player 2. We write $Q^{1}, Q^{2}$ for the probability distribution of the Nash equilibrium for players 1 and 2, respectively. For the Prisoner's Dilemma $Q^{1} = Q^{2} = (0, 1)$, i.e., the rational outcome in which no cooperation occurs.

We want to quantify the extent to which the players on the network play rationally. For this, we measure the average Jensen-Shannon divergence from the Nash equilibrium over the edges:

\begin{equation}
-\rho = \frac{1}{M}\sum_{k=1}^{M} Div(P^{1}_{k}||Q^{1}) + Div(P^{2}_{k}||Q^{2}),
\end{equation}

where

\small
\begin{equation}
Div(P||Q) = \frac{1}{2} \left( \sum_{i} P(i)\ln\frac{P(i)}{R(i)} + \sum_{i} Q(i)\ln\frac{Q(i)}{R(i)}  \right)
\label{eq:div}
\end{equation}
\normalsize

is the Jensen-Shannon divergence and $R = (P + Q)/2$  is the average of the probability distributions $P$ and $Q$ and $M$ is the number of edges in the network. We refer to $-\rho$ as the average Nash-LQRE divergence and to $\rho$ as the system rationality.

\begin{figure}
  \centering
  \begin{tabular}{@{}p{0.45\linewidth}@{\quad}p{0.45\linewidth}@{}}
    \subfigimg[width=\linewidth]{(a)}{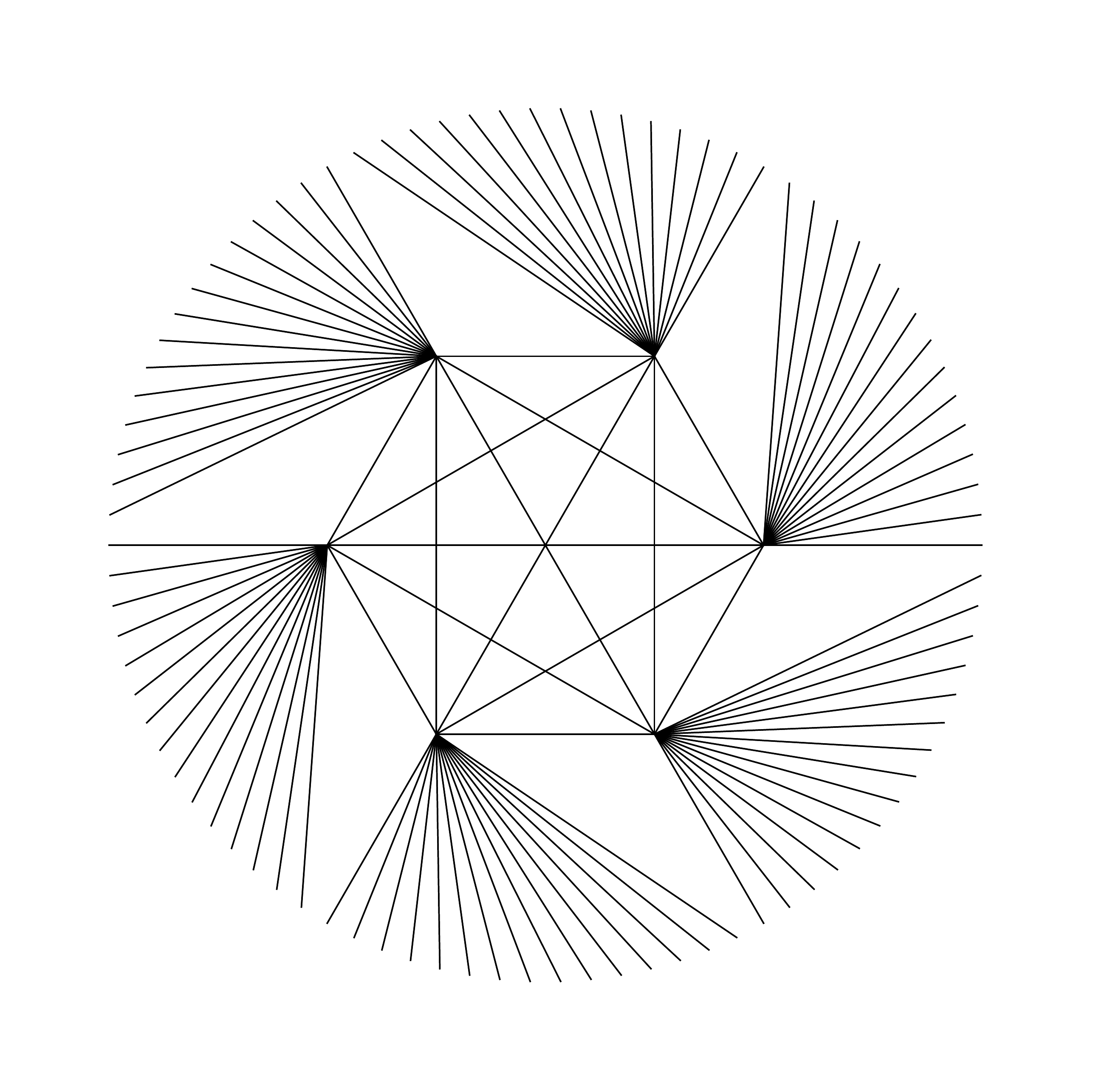} &
    \subfigimg[width=\linewidth]{(b)}{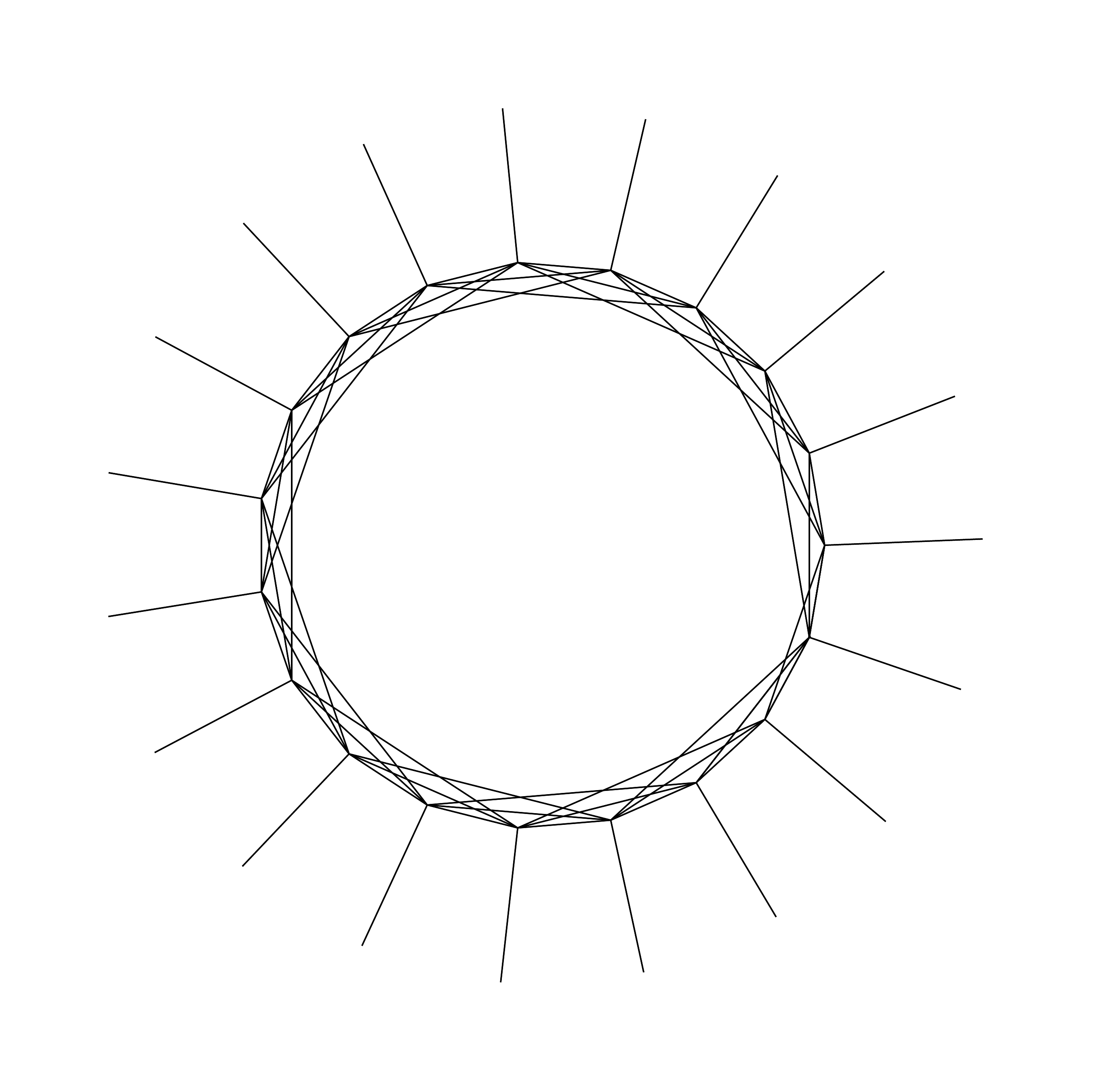}
  \end{tabular}
  \caption{Illustrative example of a core-periphery topology that minimises the average divergence (while avoiding isolated nodes) in the case of the: (a) Prisoner's Dilemma and Stag hunt games, where the hubs form a complete subgraph and each has a high degree and (b) Battle of the Sexes, where the hubs for a regular subgraph. For the Matching Pennies the optimal topology is similar to (b) but the periphery consists of isolated $K_{2}$ graphs.}
  \label{fig:CP}
\end{figure}

\subsection{Optimisation algorithm}

What is the network topology that maximises the system rationality as measured by $\rho$ or, equivalently, minimises $-\rho$? This is the question \citep{Kasthurirathna2015} posed, and the authors of \citep{Kasthurirathna2015} also suggested that scale-free graphs realise the optimum. More specifically, \citep{Kasthurirathna2015}  showed that scale-free graphs have lower $-\rho$ than random graphs with the same number of nodes and edges. Furthermore, an optimisation procedure that rewired a random graph to minimise $-\rho$ gave rise to a network with a degree distribution that fitted a power-law with a $R^{2}$ correlation of $90\%$. There are at least two weaknesses in the argument: firstly, the optimisation run by \citep{Kasthurirathna2015} does not indicate convergence to a minimum, i.e., no plateau is visibly reached and secondly, checking through linear regression that the degree distribution fits a power-law is a poor test to verify that a scale-free topology is present, and it is more robust to perform linear regression on the rank distribution \citep{Li2005}.

In the next section we propose an alternative topology that minimises $-\rho$, and give computational and analytic support for this. To numerically obtain a network with minimum $-\rho$ we consider the following numerical scheme. The scheme, which is similar to other optimisation schemes previously employed in the context of path-length optimisation \citep{Brede2010a} or optimal synchronisation \citep{Brede2010b} essentially implements a random hillclimber. Start from a random graph and repeat the following operations:
\begin{enumerate}
\item  Randomly select a node $A$, and one of its neighbors $B$.
\item The edge that links from $A$ to $B$ is rewired to connect $A$ and a randomly selected node $C$ which is not a neighbour of $A$. If we wish to prevent the appearance of isolated nodes, then we do not rewire away from nodes of degree one, i.e., we require $d(B) > 1$ when selecting $B$.
\item We calculate the divergence on each edge, according to \eqref{eq:div}, in the new neighborhood of $C$ and sum them. We compare this sum to the analogous sum computed for the old neighborhood of $B$.
\item  If there is a decrease in the value of the sum, the rewiring is kept. Otherwise, the original position of the edge is restored and the entire procedure is repeated.
\item If we allow isolated nodes to form during the optimisation, then we regularly remove them and randomly rewire the links between the remaining nodes (akin to simulated annealing). Afterwards, the optimisation proceeds as above. 
\end{enumerate}

The above algorithm implies that, after a re-wiring, the rationalities of a node and its neighbours are updated simultaneously. Note that, once a node becomes isolated, it will never reconnect to the rest of network through the hill-climbing part of the optimisation. This is because a zero degree node will always increase the average divergence if reconnected to the rest of graph through an edge.

In the following section we reproduce the results of \citep{Kasthurirathna2015} for random and scale-free networks, and also consider two additional topologies, regular random graphs and what we refer to as core-periphery networks. The core-periphery network we explore can be described as having a bipartite subgraph, but, in addition, the high degree nodes, i.e., hubs, form  a regular (or even complete) subgraph. See Fig. \ref{fig:CP} for a  illustration of the types of core-periphery networks we find when maximising system rationality.
\section{Results}
\label{sec:rs}

\bgroup
\def\arraystretch{1.5}% 
\begin{table*}[t]
\centering
\begin{tabular}{|c|c|c|c|c|c|c|c|c|c|c|}
\hline
\begin{tabular}[c]{@{}c@{}}Game with\\ $\beta  = 1.33$\end{tabular} & \multicolumn{5}{c|}{Prisoner's Dilemma} & \multicolumn{5}{c|}{Stag Hunt} \\ \hline
\multirow{2}{*}{Graph type} & \multicolumn{2}{c|}{Core-periphery ($K=13$)} & \multirow{2}{*}{Scale-free} & \multirow{2}{*}{Random} & \multirow{2}{*}{Regular} & \multicolumn{2}{c|}{Core-periphery ($K=18$)} & \multirow{2}{*}{Scale-free} & \multirow{2}{*}{Random} & \multirow{2}{*}{Regular} \\ \cline{2-3} \cline{7-8}
 & Computed & Analytic &  &  &  & Computed & Analytic &  &  &  \\ \hline
Linear & 0.2149(2) & 0.2135 & 0.361(2) & 0.3914(3) & 0.3982 & 0.1532(6) & 0.1485 & 0.337(7) & 0.311(1) & 0.3377 \\ \hline
Convex & 0.2112(2) & 0.2108 & 0.407(4) & 0.4290(1) & 0.4300 & 0.1988(1) & 0.1983 & 0.395(4) & 0.4263(1) & 0.4286 \\ \hline
Concave & 0.3082(1) & 0.2890 & 0.377(1) & 0.3876(2) & 0.3907 & 0.2435(5) & 0.3817 & 0.284(3)  & 0.2930(7) & 0.3067 \\ \hline
\end{tabular}
\caption{The average Nash-LQRE divergence ($-\rho$) in the case of the Prisoner's Dilemma and Stag Hunt games for the different network topologies and rationality functions. The numerical results for the core-periphery, scale-free and random networks are averaged over 100 instances. The core-periphery nodes have $K$ hubs.}
\label{table:PDSHdiv}
\end{table*}
\egroup

We analyse four games: the Prisoner's Dilemma, the Stag Hunt, the Battle of the Sexes and the Matching Pennies game. The network topologies we investigate are: core-periphery, scale-free networks, regular and Erd\H{o}s-R\'{e}nyi random graphs. The scale-free networks are generated using preferential attachment, according to the Barab\'{a}si-Albert model \citep{Barabasi1999}. Tables \ref{table:PDSHdiv} and \ref{table:BSMPdiv} show the average divergence in the case of the four games on networks that have $N = 1000$ nodes and $M = 2000$ edges. The results for the random and scale-free graphs are averaged over 100 instances, and likewise for the core-periphery topology in the case of the Prisoner's Dilemma and Stag Hunt games. The `game parameter' is set to $\beta = 1.33$. The choice of games, network parameters and $\beta$ parameter are again consistent with \citep{Kasthurirathna2015}. We deviate from \citep{Kasthurirathna2015} by considering a more common form of the Matching Pennies that is employed in experimental setups \citep{Ochs1995}. In the case of the asymmetric games, namely the Battles of the Sexes and Matching Pennies, the game on each edge is played twice with players interchanging roles. This amounts to solving equations \eqref{eq:qre} once with rationalities $(\lambda_{1}, \lambda_{2})$ and once with $(\lambda_{2}, \lambda_{1})$, and then taking the average of the two divergences.

\bgroup
\def\arraystretch{1.5}% 
\begin{table}[t]
\centering
\begin{tabular}{|c|c|c|llclll|c|c|c|}
\cline{1-3} \cline{10-12}
PD & \multicolumn{2}{c|}{P1} &  &  &  &  &  &  & SH & \multicolumn{2}{c|}{P1} \\ \cline{1-3} \cline{10-12} 
\multirow{2}{*}{P2} & 1, 1 & 0, $\beta$ &  &  &  &  &  &  & \multirow{2}{*}{P2} & $\beta, \beta$ & 0, 1 \\ \cline{2-3} \cline{11-12} 
 & $\beta$, 0 & 0, 0 &  &  &  &  &  &  &  & 1, 0 & 0, 0 \\ \cline{1-3} \cline{10-12} 
\end{tabular}
\caption{Payoffs for the Prisoner's Dilemma (PD) and Stag Hunt (SH) games.}
\label{table:PDSHpayoff}
\end{table}
\egroup

\subsection{Symmetric games}
\label{subsec:pdsh}

\subsubsection{Prisoner's Dilemma}

\begin{figure*}[t]
  \centering
  \begin{tabular}{@{}p{0.5\linewidth}@{\quad}p{0.5\linewidth}@{}}
    \subfigimg[width=\linewidth]{(a)}{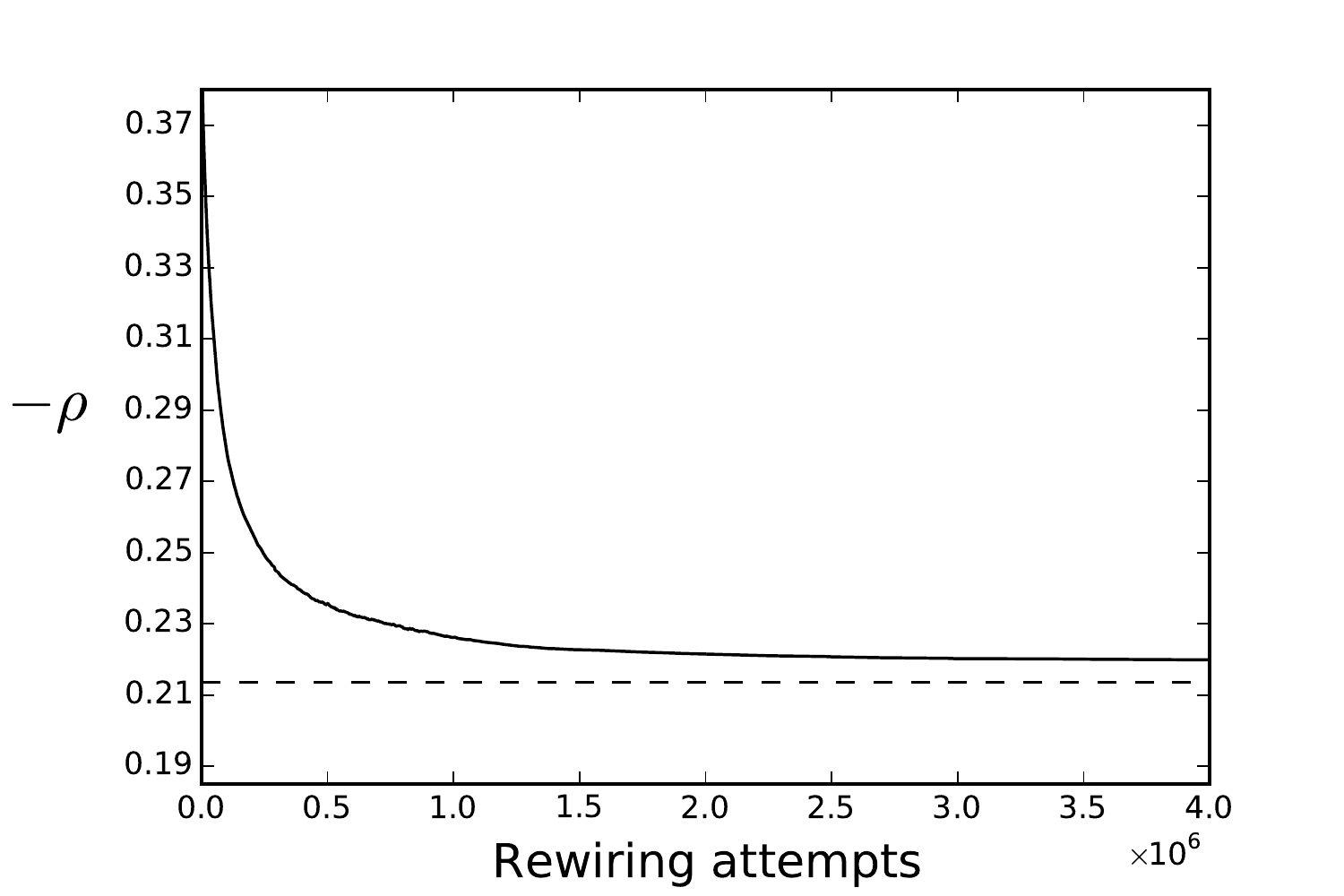} &
    \subfigimg[width=\linewidth]{(b)}{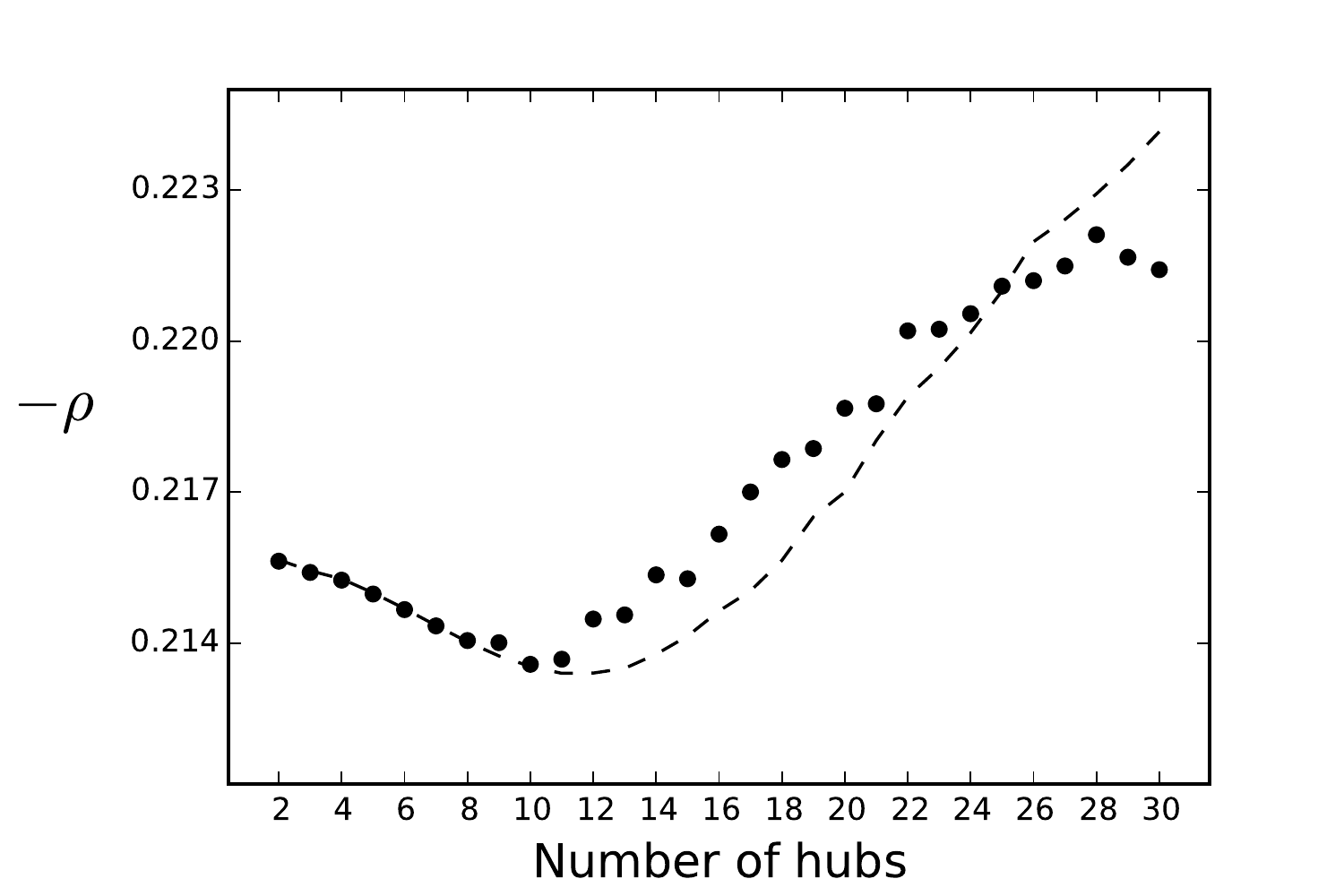}
  \end{tabular}
\caption{(a) The evolution of $-\rho$, the average Jensen-Shannon divergence between Nash and LQR equilibria, for the Prisoner's Dilemma ($\beta = 1.33$) with linear rationality, of an initially random network ($N = 1000$ nodes and $M = 2000$ edges) as a function of attempted edge rewirings, while not allowing isolated nodes. The optimisation gets within $2\%$ of the estimated minimum (dashed line). (b) The divergence of a core-periphery graph with $N=1000, M=2000$ as a function of the number of hubs according to numerical calculations (dots) and analytic estimates (dashed line) using equation \eqref{eq:div_cp} for core-periphery graphs.}
\label{fig:PD1}
\end{figure*}

Our results for the Prisoner's Dilemma on random networks closely match with \citep{Kasthurirathna2015}, while the results for scale-free networks are consistent but show a larger decrease than reported in \cite{Kasthurirathna2015} compared to the random case, see Table \ref{table:PDSHdiv}.
Next we consider rationality-optimised networks. Figure \ref{fig:PD1}(a) shows the evolution of the divergence during optimisation. The numerical computations indicate that the topology of the random graph is evolving towards a core-periphery graph with a complete subgraph of $K = 13$ hubs, cf. Fig. \ref{fig:CP} for a qualitative visualisation. To complement these simulations with analytic arguments we approximate the solutions to equations \eqref{eq:qre}.

The network obtained through the optimisation procedure has a divergence within $2\%$ of the analytic estimate, as Fig. \ref{fig:PD1}(a) shows. The core-periphery networks that were generated to compute the entries in Table \ref{table:PDSHdiv} have a divergence that deviates less than $0.5\%$ from the analytic estimate (in the case of linear rationality). Hence, the difference between the numerical result and the analytic estimate in Fig. \ref{fig:PD1}(a) likely stems from the optimisation getting stuck in a local optimum. Selective re-wirings of a few edges could further reduce the divergence but the global network topology would still be of core-periphery type with $K = 13$ hubs.

The payoffs for the Prisoner's Dilemma are shown in Table \ref{table:PDSHpayoff}. In this case we have that $a_{1} = 1, b_{1}= e^{\lambda_{1}(\beta-1)}, a_{2} = 1, b_{2} = e^{\lambda_{2}(\beta-1)}$ and equations \eqref{eq:qre} take the form:
\begin{equation}
\begin{aligned}
p^{1}_{c}(1+e^{\lambda_{1}(\beta-1)p^{2}_{c}}) &= 1\\
p^{2}_{c}(1+e^{\lambda_{2}(\beta-1)p^{1}_{c}}) &= 1
\end{aligned}
\end{equation}

We can obtain analytic results for regular graphs in which all players are characterised by the same rationality. Let $p_{r}$ be the probability of cooperation on a regular graph. Then, in the case of the Prisoner's Dilemma,  $p_{r}$ satisfies the equation:
\begin{equation}
p_{r}(1+ e^{\lambda (\beta - 1) p_{r}}) = 1,
\end{equation}
where $\lambda$ is the rationality of a node. The divergence $-\rho_{r}$ for a regular graph with $N$ nodes and $M$ edges is thus given by:
\begin{equation}
-\rho_{r} = p_{r} \ln{2} + (1-p_{r}) \ln{\frac{1-p_{r}}{1-p_{r}/2}} + \ln{\frac{1}{1-p_{r}/2}}.
\label{eq:div_reg}
\end{equation}

The results of applying equation \eqref{eq:div_reg} for the different rationality functions match precisely with the entries for regular random graphs in Table \ref{table:PDSHdiv}. The degree distribution for the random graph has an average and mode of $\langle d \rangle = 4$, and we see that results for the random and regular topologies are close (within $2\%$), i.e., the regular random graph can be seen as a zero order approximation of the random graph.

Analytic results are also possible for the core-periphery graph, where the core consists of hubs with high degree (high rationality) and the periphery contains nodes of low degree (low rationality).  Let $p_{h}$ be the probability of cooperation for a hub and $p_{f}$ the same quantity but for a periphery (fringe) node. For the Prisoner's Dilemma, the equations $p_{h}$ and $p_{f}$ satisfy are
\begin{equation}
\begin{aligned}
p_{h}(1 + e^{\lambda_{h} p_{f} (\beta - 1)}) &= 1\\
p_{f}(1 + e^{\lambda_{f} p_{h} (\beta - 1)}) &= 1
\end{aligned}
\end{equation}

where $\lambda_{h}$ and $\lambda_{f}$ are the rationalities of the two types of nodes. If $\lambda_{f} > 0$ and $\lambda_{h} \gg 1$, then $p_{h} \simeq 0$ and this implies that $p_{f} = 0.5$. So, we can approximate the probability of cooperation of a periphery node with a hub to zero order by $p^{(0)}_{f} = 1/2$ , while hubs do not cooperate, i.e., $p^{(0)}_{h} = 0$. The contribution of the edges between the high connectivity nodes in a core-periphery graph leaves the divergence -$\rho$ largely unchanged, because the hubs are highly rational (due to their large degree). The divergence of a connected core-periphery graph with a finite number of hubs but infinite number of periphery nodes is thus given by $\mu = 3/4\ln 4/3 = 0.216$, independent of the choice of rationality function is long as it is an increasing function of the degree.

For a more accurate picture we consider first order corrections and obtain
\begin{equation}
\begin{aligned}
p_{h} &= \frac{1}{1+e^{\lambda_{h}(\beta-1)/2}} \simeq e^{-\lambda_{h}(\beta-1)/2} = p_{h}^{(1)}\\
p_{f} &= \frac{1}{1+e^{\lambda_{f}(\beta-1)p^{1}_{c}}} \simeq \frac{1}{2(1+\lambda_{f}(\beta-1)p^{1}_{c}/2)}\\
&\simeq \frac{1}{2}-\frac{\lambda_{f}(\beta-1)p^{1}_{c}}{4} = p_{f}^{(0)} + p_{f}^{(1)}
\end{aligned}
\label{eq:pd1st}
\end{equation}

In a core-periphery network with $K$ hubs each each hub has degree $M/K + K-1$ and the periphery nodes have average degree $d$ that satisfies
\begin{equation}
\frac{K(K-1)}{2}+d(N-K)=M.
\end{equation}

If the hubs are of high enough degree such that the first order corrections $p_{h}^{(1)}, p_{f}^{(1)}$ are small, then the divergence -$\rho$ of the core-periphery graph with $N$ nodes and $K$ hubs can be approximated as follows:
\begin{equation}
\begin{aligned}
-\rho_{cp} \simeq &\left(1-\frac{K(K-1)}{2M}\right)\left(\mu + \frac{p_{h}^{(1)}\ln{2}+p_{f}^{(1)}\ln{3}}{2}\right)\\ &- \frac{K(K-1)}{2M}\rho_{c},
\end{aligned}
\label{eq:div_cp}
\end{equation}
where $-\rho_{c}$ is the average divergence over the edges between the hubs, which form a complete subgraph. The average divergence of the core-periphery network with $K = 13$ hubs for linear rationality is $-\rho_{cp} = 0.2135$. This result is close but slightly lower than the numerically computed entries in Table \ref{table:PDSHdiv}. As one would expect, the approximation is better for the convex rationality function (because it extends the region in which the approximation \eqref{eq:pd1st} holds), whereas the analytic estimate is poor for the concave case (due to the opposite effect).

\begin{figure*}[t]
  \centering
  \begin{tabular}{@{}p{0.5\linewidth}@{\quad}p{0.5\linewidth}@{}}
    \subfigimg[width=\linewidth]{(a)}{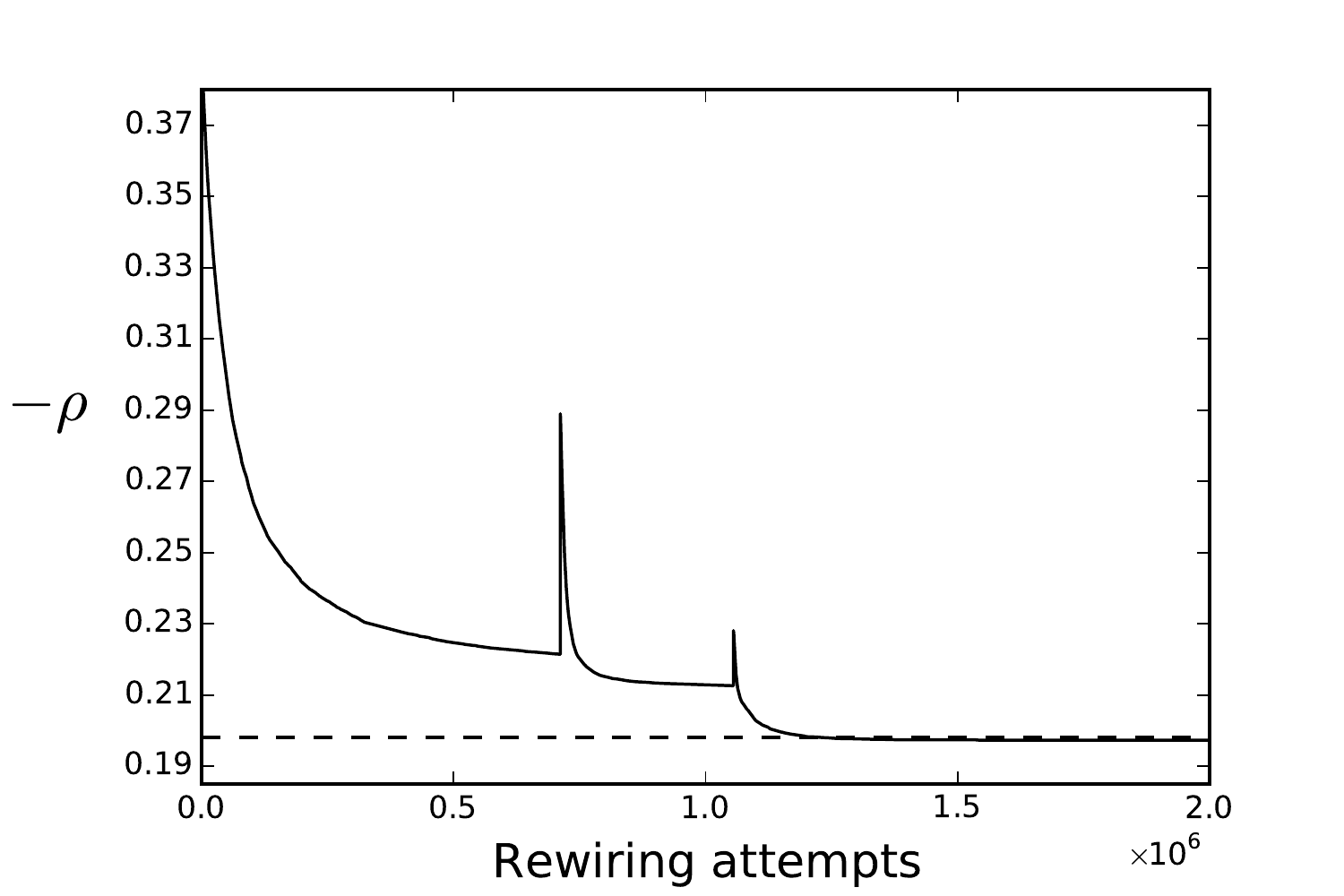} &
    \subfigimg[width=\linewidth]{(b)}{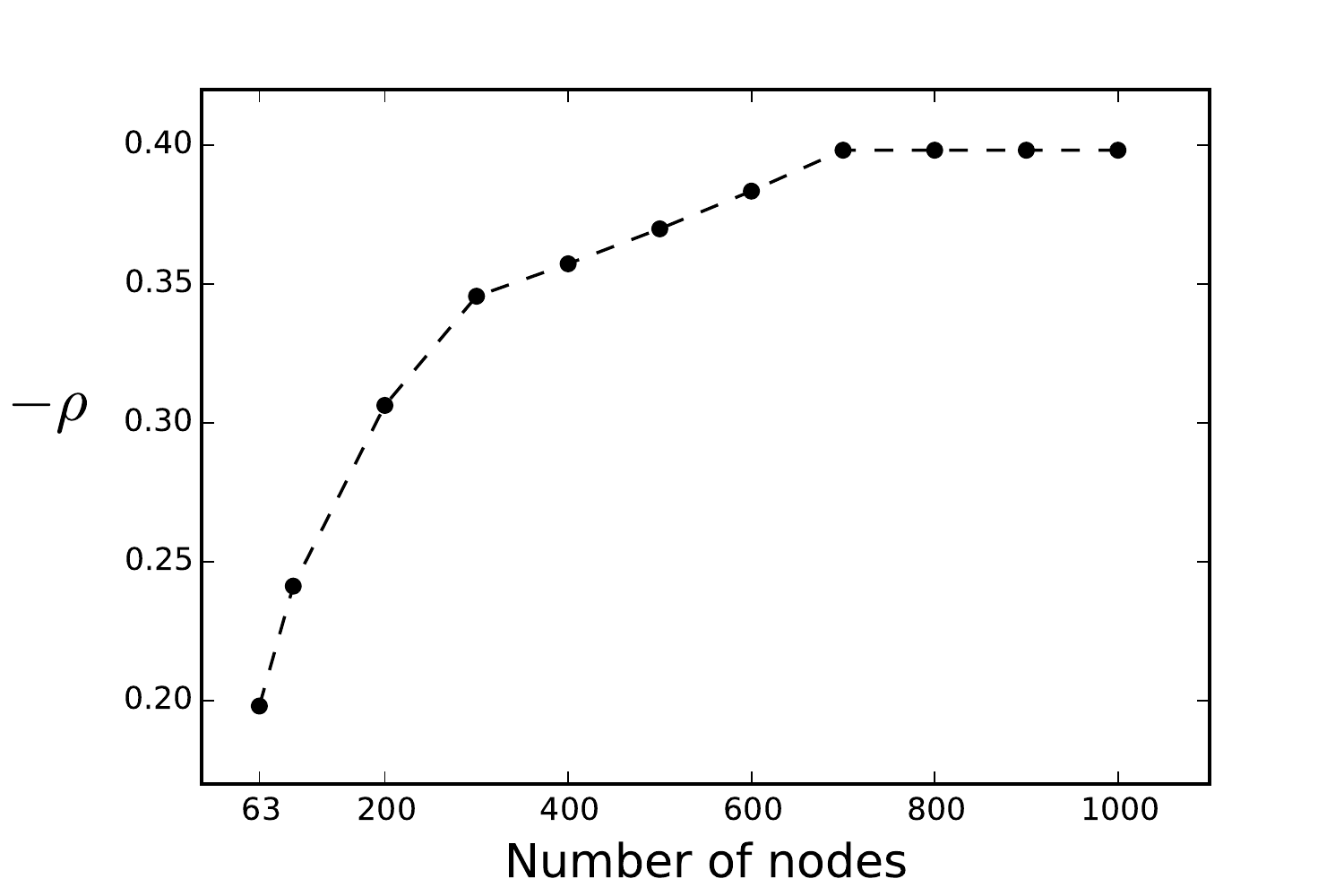}
  \end{tabular}
\caption{(a)  The evolution of $-\rho$, the average Jensen-Shannon divergence between Nash and LQR equilibria, for the Prisoner's Dilemma ($\beta = 1.33$) with linear rationality, starting with an initially random network ($N = 1000$ nodes and $M = 2000$ edges) as a function of the number of attempted edge rewirings, allowing isolated nodes. The optimisation reaches the theoretical minimum (dashed line), which is achieved for a complete graph with 63 nodes. (b) Numerical (dots) and analytical (dashed line) results for the divergence of regular graphs with $M=2000$ edges in the case of the Prisoner's Dilemma. Analytic values calculated according to equation \eqref{eq:div_reg}.} 
\label{fig:PD2}
\end{figure*}

Making use of Eq. \eqref{eq:div_cp} Fig. \ref{fig:PD1}(b) shows the dependence of the Nash-LQRE divergence as a function of the number of hubs in the core-periphery network. We can understand the dependence by noting the two competing factors in equation \eqref{eq:div_cp}. With increasing numbers of hubs, the number of edges in the core increases which tends to lower the average divergence. On the other hand in a finite graph, an increasing number of hubs leads to a decrease of the average degree of a hub, which makes hubs less rational, and so the divergence tends to increase. For a graph with $N=1000$ nodes and $M=2000$ edges the minimum of \eqref{eq:div_cp} is obtained in the neighbourhood of 12 hubs, in good agreement with the numerical optimisation in Fig. \ref{fig:PD1}(a) that tends toward a graph with 13 hubs.

As Table \ref{table:PDSHdiv} shows, the results for the average divergence of the core-periphery graph are notably smaller than results obtained for all other topologies. Furthermore, it is easy to see that the core-periphery networks lie at local minima (in the space of possible graphs) with respect to the value of the Nash-LQRE divergence. This can be seen by exploring the effects of possible rewirings. Moving an edge that connects two nodes in the core (a $cc$ edge) to connect another two nodes in core does not change the average divergence. Also, moving an edge that connects a core node to a periphery node (a $cp$ edge) to connect another core-periphery pair has no effect. A rewiring that does affect the divergence involves moving an edge that lies between:
\begin{enumerate}[label=(\alph*), nolistsep]
\item  two hubs, so that it connects a hub and a periphery node, i.e., moving from a $cc$ edge to a $cp$ edge
\item two hubs, so that it connects two periphery nodes, i.e., moving from a $cc$ edge to a $pp$ edge
\item a hub and a periphery node, so that it connects two periphery nodes, i.e., moving from a $cp$ edge to a $pp$ edge
\end{enumerate}

Due to the high rationality of hub nodes, an edge between two hubs has a near zero contribution to the divergence. The low degree nodes are associated with low rationality, hence they will a have quantal response equilibria that differ significantly from the Nash equilibrium. Hence, moving an edge that lies between two hubs so that it now connects to one or two periphery nodes will lead to an increase the in the average divergence.

Moving an edge that connects a hub and a periphery node to connect two periphery nodes, will also lead to an increase in the divergence, which we can quantify. After rewiring an edge that connected a hub to the periphery to now connect two periphery nodes, the divergence is:

\small
\begin{equation}
\begin{aligned}
-\rho_{w} \simeq &\left(1-\frac{K(K-1)/2-1}{M}\right)\left(\mu + \frac{p_{h}^{(1)}\ln{2}+p_{f}^{(1)}\ln{3}}{2}\right) \\&- \frac{K(K-1)}{2M}\rho_{c}
+ \frac{1}{M} p_{d+1} \ln{2} \\&+ \frac{1}{M}  \left((1-p_{d+1}) \ln{\frac{1-p_{d+1}}{1-p_{d+1}/2}} + \ln{\frac{1}{1-p_{d+1}/2}}\right)
\end{aligned}
\end{equation}
\normalsize

where $p_{d+1}$ is the probability of cooperation for a node in a regular graph of average degree $d+1$. If $p_{d+1}\simeq 0.5$, then the last three terms are approximately $2\mu/M$, which is larger than the term it is replacing that is approximately $\mu/M$. Thus, $-\rho_{w} \geq -\rho_{cp} $. In summary, any rewiring either does not change the average divergence, or increases it. So, the core-periphery topology achieves a local minimum for the average divergence in the space of possible graph topologies.

\bgroup
\def\arraystretch{1.5}% 
\begin{table*}[t]
\centering
\begin{tabular}{|c|c|l|c|c|c|c|l|c|c|c|}
\hline
\begin{tabular}[c]{@{}c@{}}Game with\\ $\beta  = 1.33$\end{tabular} & \multicolumn{5}{c|}{Battle of the Sexes ($-\rho \times 10^{2}$)} & \multicolumn{5}{c|}{Matching Pennies  ($-\rho \times 10^{2}$)} \\ \hline
\multirow{2}{*}{Graph type} & \multicolumn{2}{c|}{\multirow{2}{*}{$\begin{matrix} \text{Core-periphery} \\ \text{(K=1500)} \end{matrix}$}} & \multirow{2}{*}{Scale-free} & \multirow{2}{*}{Random} & \multirow{2}{*}{Regular} & \multicolumn{2}{c|}{\multirow{2}{*}{$\begin{matrix} \text{Core-periphery} \\ \text{(K=168)} \end{matrix}$}} & \multirow{2}{*}{Scale-free} & \multirow{2}{*}{Random} & \multirow{2}{*}{Regular} \\
 & \multicolumn{2}{c|}{} &  &  &  & \multicolumn{2}{c|}{} &  &  &  \\ \hline
Linear & \multicolumn{2}{c|}{0.178} & 2.9(3) & 0.25(1) & 0.236 & \multicolumn{2}{c|}{0.096} & 0.47(3) & 0.215(1) & 0.218 \\ \hline
Convex & \multicolumn{2}{c|}{0.458} & 1.0(4) & 0.473(1) & 0.486 & \multicolumn{2}{c|}{0.230} & 0.8(3) & 0.252(1) & 0.252 \\ \hline
Concave & \multicolumn{2}{c|}{0.180} & 1.1(1) & 0.200(1) & 0.202 & \multicolumn{2}{c|}{0.142} & 0.26(1) & 0.200(1) & 0.203 \\ \hline
\end{tabular}
\caption{The average Nash-LQRE divergence ($-\rho$) in the case of the Battle of the Sexes and Matching Pennies games for the different network topologies and different choices of rationality functions. The numerical results for the scale-free and random networks are averaged over 100 instances. The core-periphery nodes have $K$ hubs.}
\label{table:BSMPdiv}
\end{table*}
\egroup

\subsubsection{Stag Hunt}

The results for the Stag Hunt game are largely analogous to the Prisoner's Dilemma. The payoff matrix for the Stag Hunt is also in Table \ref{table:PDSHpayoff}. We have that $a_{1} = e^{\lambda_{1}}, b_{1}= e^{-\lambda_{1}\beta}, a_{2} = e^{\lambda_{2}}, b_{2} = e^{-\lambda_{2}\beta}$ and equations \eqref{eq:qre} take the form:
\begin{equation}
\begin{aligned}
p^{1}_{c}(1+e^{\lambda_{1}(1-\beta p^{2}_{c})}) &= 1\\
p^{2}_{c}(1+e^{\lambda_{2}(1-\beta p^{1}_{c})}) &= 1
\end{aligned}
\end{equation}

For a regular graph the equations reduce to:
\begin{equation}
p_{r}(1+e^{\lambda (1-\beta p_{r})}) = 1
\label{eq:reg_sh}
\end{equation}

Applying Eq. \eqref{eq:div_reg} for the average divergence of a regular graph to the solution of \eqref{eq:reg_sh} again matches with the numerical computations in Table \ref{table:PDSHdiv}.

If $\lambda_{h}$ is large and $\lambda_{f}$ is small then we can approximate to zero order $p_{h} = 0$ and $p_{f} = 1/(1+e^{\lambda_{f}})$. If  $\beta < 1 + e^{\lambda_{2}}$, then the first order approximations are: 
\begin{equation}
\begin{aligned}
p_{h} &= \frac{1}{1+e^{( \lambda_{h} (1-\beta/(1+e^{\lambda_{f}}) ))}} \simeq  e^{-\lambda_{1}  (1-\beta/(1+e^{\lambda_{f}}) )}\\
p_{f} &= \frac{1}{1+e^{\lambda_{f}(1 - \beta p_{h})}} \simeq \frac{1}{2(1-\lambda_{f}(1 - \beta p_{h})/2)}\\
&\simeq \frac{1}{2} - \frac{\lambda_{f}(1 - \beta p_{h})}{4}
\end{aligned}
\label{eq:sh1st}
\end{equation}

The optimum network that minimises $-\rho$ is a core-periphery network with $K = 18$ hubs that form a complete subgraph. For linear rationality, equation \eqref{eq:div_cp} gives an average divergence of $-\rho_{cp} = 0.1485$ which is in reasonable agreement with the numerical calculation in Table \ref{table:PDSHdiv}, while a close match is found for convex rationality. A large difference is observed between the numerical result and analytic estimate for the concave case, due to the square root function decreasing the rationality of the hubs and, thus, approximation \eqref{eq:sh1st} failing.

Above we have considered rewirings in which isolated nodes were prevented. However, the observation of the emergence of a strong core-periphery structure makes one wonder if optimal networks would actually become disconnected when this constraint is removed. Hence, we consider again the Prisoner's Dilemma and relax the optimisation conditions in the rewiring of the network to allow for isolated nodes. Figure \ref{fig:PD2}(a) shows the results of a variant of simulated annealing performed to optimise the network. The numerical results indicate that a majority of nodes become isolated and a certain number of hub nodes end up sharing the edges between them. If we approximate the remaining sub-graph of hubs as a complete graph, then the number of hubs is given by
\begin{equation}
K = \frac{1+\sqrt{1+8M}}{2} 
\end{equation}
The average divergence for a complete graph with $K$ nodes is given by equation \eqref{eq:div_reg} for a regular graph when the average degree is $K$. For $M = 2000$ the number of hubs is $K= 63$ and gives a divergence of $-\rho = 0.197$, which matches with the numerical optimisation in Fig. \ref{fig:PD2}(a). Of all regular graphs with $M = 2000$ edges the complete graph with $63$ nodes has minimum divergence, which is indicated in Fig. \ref{fig:PD2}(b). For the Stag Hunt game the unconstrained optimisation leads to regular graph with $N=400$ nodes and average degree $2M/N = 10$.

\subsection{Asymmetric games}
\label{subsec:bsmp}

\bgroup
\def\arraystretch{1.5}% 
\begin{table}[t]
\centering
\begin{tabular}{|c|c|c|llclll|c|c|c|}
\cline{1-3} \cline{10-12}
BS & \multicolumn{2}{c|}{P1} &  &  &  &  &  &  & MP & \multicolumn{2}{c|}{P1} \\ \cline{1-3} \cline{10-12} 
\multirow{2}{*}{P2} & $\beta$, 1 & 0, 0 &  &  &  &  &  &  & \multirow{2}{*}{P2} & $\beta$, 0 & 0, 1 \\ \cline{2-3} \cline{11-12} 
 & 0, 0 & 1, $\beta$ &  &  &  &  &  &  &  & 0, 1 & 1, 0 \\ \cline{1-3} \cline{10-12} 
\end{tabular}
\caption{Payoffs for the Battle of the Sexes (BS) and Matching Pennies (MP) games.}
\label{table:BSMPpayoff}
\end{table}
\egroup

\begin{figure*}[t]
  \centering
  \begin{tabular}{@{}p{0.5\linewidth}@{\quad}p{0.5\linewidth}@{}}
    \subfigimg[width=\linewidth]{(a)}{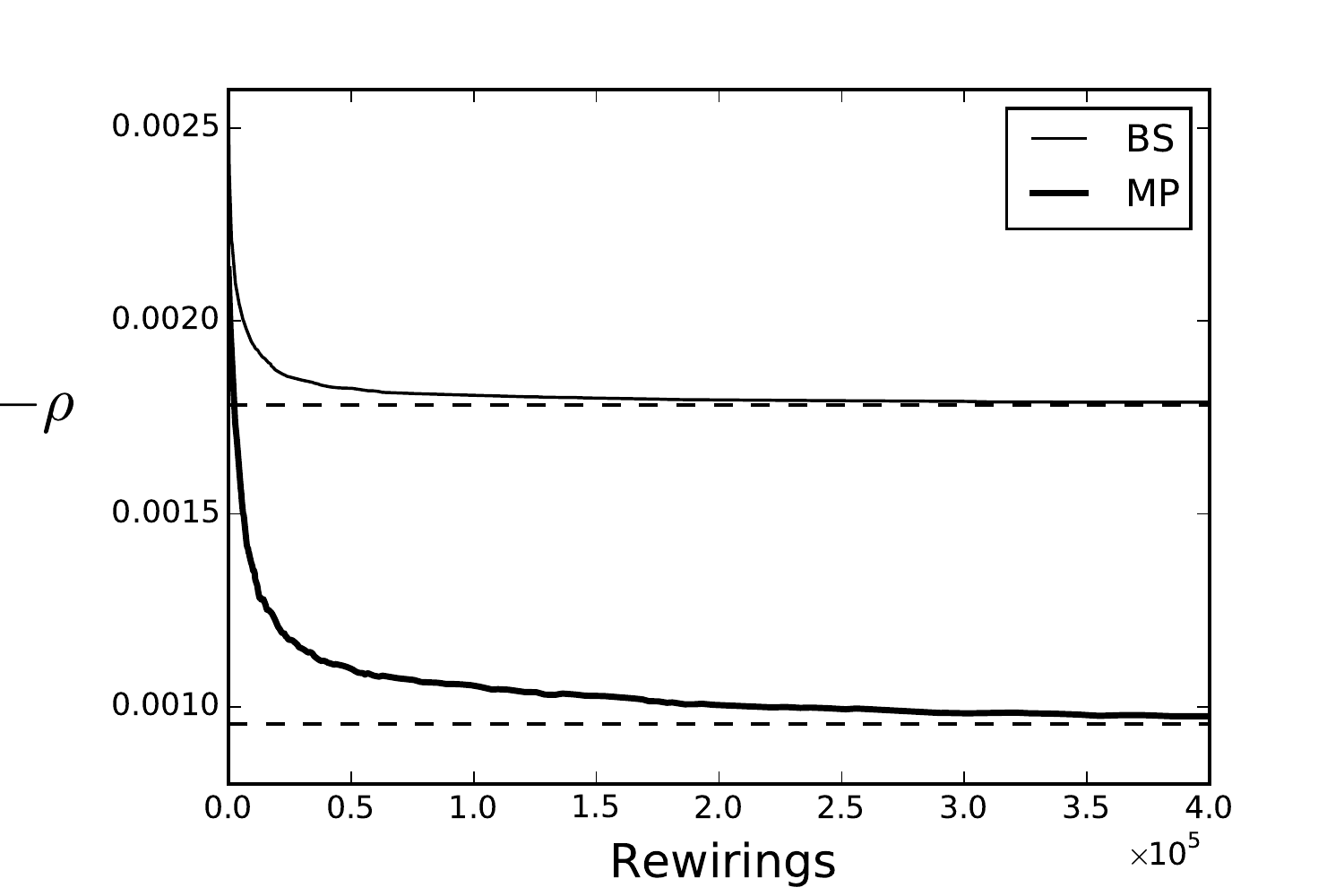} &
    \subfigimg[width=\linewidth]{(b)}{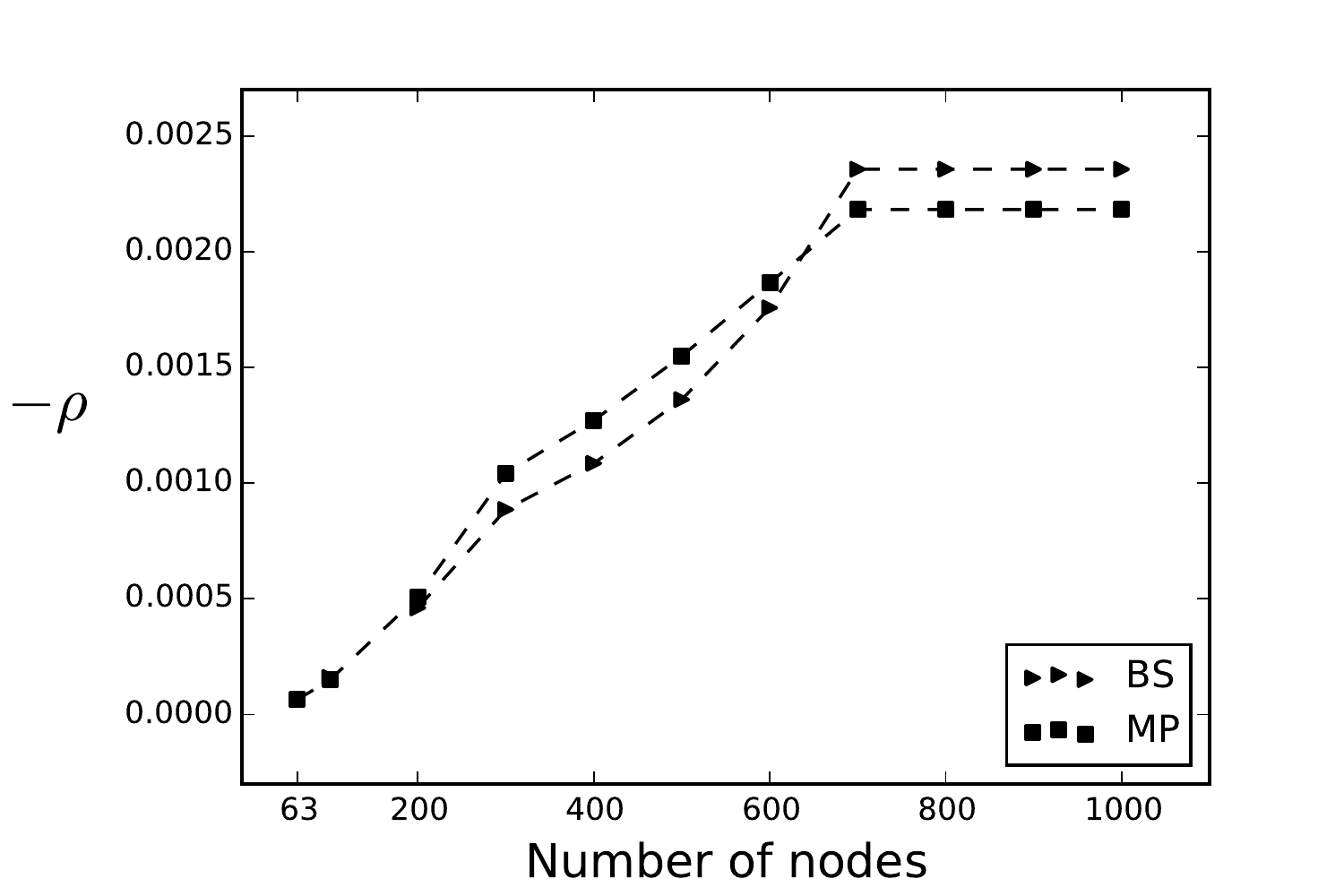}
  \end{tabular}
\caption{(a) The evolution of $-\rho$, the average Jensen-Shannon divergence between Nash and LQR equilibria, for the Battles of the Sexes (thin line) and Matching Pennies (thick line) ($\beta = 1.33$) with linear rationality, in the range of topologies between an initially random network ($N = 1000$ nodes and $M = 2000$ edges) as a function of attempted edge rewirings, while not allowing isolated nodes. The optimisations reaches the estimated minima (dashed lines). (b) The divergence for regular graphs with $M=2000$ edges in the case of the Battle of the Sexes (triangular dots) and Matching Pennies (square dots), with analytic results shown by dashed lines.}
\label{fig:BSMPdiv}
\end{figure*}

\subsubsection{Battle of the Sexes}

We now consider the asymmetric games: the Battle of the Sexes and the Matching Pennies, with payoffs given in Table \ref{table:BSMPpayoff}. In the cases of these two games we compute the average Nash-LQRE divergence with respect to the mixed strategy Nash equilibria. For the Battle of the Sexes, we find that the average divergence with respect to the mixed Nash equilibrium, $Q^{1} = (\beta/(1+\beta),1/(1+\beta))$ and $Q^{2} = 1-Q^{1}$, is much closer to zero than for the pure strategy $Q^{1} = Q^{2} = (0, 1)$.  In the case of the Matching Pennies games, there is no pure Nash equilibrium.

We first analyse the Battle of the Sexes for which $a_{1} = e^{\lambda_{1}}, b_{1}= e^{-\lambda_{1}(1+\beta)}, a_{2} = e^{\lambda_{2}\beta}, b_{2} = e^{-\lambda_{2}(1+\beta)}$.
Equations \eqref{eq:qre} take the form:

\begin{equation}
\begin{aligned}
p^{1}_{c}(1+e^{\lambda_{1}(1- (1 + \beta) p^{2}_{c})}) &= 1\\
p^{2}_{c}(1+e^{\lambda_{2}(\beta-(1+\beta) p^{1}_{c})}) &= 1
\end{aligned}
\label{eq:bs}
\end{equation}

As equations \eqref{eq:bs} are not symmetric when exchanging players, the game is played twice on each edge, once solving equations \eqref{eq:qre} with the rationalities $(\lambda_{1}, \lambda_{2})$, and once with the rationalities interchanged, i.e, $(\lambda_{2}, \lambda_{1})$.

If $\lambda_{1}$ and $\lambda_{2}$ are small then we can approximate to zero order and obtain $p^{1}_{c} = 1/2$ and $p^{2}_{c} = 1/2$. The approximations to first order are given by:

\begin{equation}
\begin{aligned}
p^{1}_{c} &= \frac{1}{1+e^{ \frac{\lambda_{1}(1-\beta)}{2}}} \simeq \frac{1}{2\left(1+\frac{\lambda_{1}(1-\beta)}{4}\right)} \simeq \frac{1}{2}+\frac{\lambda_{1}(\beta-1)}{8}\\
p^{2}_{c} &= \frac{1}{1+e^{ \frac{\lambda_{2}(\beta-1)}{2}}} \simeq \frac{1}{2\left(1+\frac{\lambda_{2}(\beta-1)}{4}\right)} \simeq \frac{1}{2}-\frac{\lambda_{2}(\beta-1)}{8}
\end{aligned}
\label{eq:bs1st}
\end{equation}

The average divergence for different topologies is found in Table \ref{table:BSMPdiv}. By optimising the network topology to minimise $-\rho$, as we did for the Prisoner's Dilemma game, we obtain the results in Fig. \ref{fig:BSMPdiv}. The divergence is minimised for a core-periphery network where the core consist of $N/2 = 500$ nodes that form a regular graph, and each node in the core links to a single periphery node. The degree of a node in the core is $d_{h} = 7$, while in the periphery it is $d_{f} = 1$. A qualitative illustration of the network topology can be seen in Fig. \ref{fig:CP}(b). 

If we optimise the network topology to minimise the divergence with respect to the pure strategy Nash equilibrium $Q^{1} = Q^{2} = (0, 1)$, we obtain a regular graph of average degree $2M/N = 4$ and $-\rho = 2\mu = 0.432$ for all rationality functions. As Table \ref{table:BSMPdiv} shows, optimising with respect to the mixed strategy Nash equilibrium gives a much lower value for the average divergence $-\rho$, which is why we focus on this case.

We can gain analytic insight into the results presented in Table \ref{table:BSMPdiv} by writing the average divergence as:
\begin{equation}
-\rho = \alpha (-\rho_{c}) + (1-\alpha)(-\rho_{f})
\label{eq:div_asym}
\end{equation}
where $-\rho_{c}$ is the average divergence of the core, i.e., that of a regular graph with average degree $d_{h} = 7$, and $-\rho_{f}$ is the average divergence of the periphery, i.e., the divergence for an edge connecting a degree $d_{f} = 1$ node with a node from the core. The value of $\alpha$ is the fraction of links that exist within the subgraph containing the core. Hence, $\alpha = N(d_{h}-1)/(2M) = 75\%$. The rest of the links are part of the periphery. By numerically solving for the cooperation probabilities in \eqref{eq:bs} and using equation \eqref{eq:div_asym}, we obtain an exact match to the numerically obtained entries in Table \ref{table:BSMPdiv}.

If we allow isolated nodes to form throughout the optimisation then, similar to the Prisoner's Dilemma and Stag Hunt games, the periphery nodes disconnect from the network but now a core with a strongly bi-modal degree distribution emerges. The core has $N_{c1} = 263$ nodes with degree $d_{h1} \simeq 8$, and $N_{c2} = 159$ nodes with degree $d_{h2} \simeq 12$. The average divergence is approximated well by \eqref{eq:div_asym}, where the two contributions come from regular graphs with $N_{c1}, N_{c2}$ nodes with average degrees $8$ and, respectively, $12$.  Numerical and analytic results for regular graphs for the asymmetric games we consider are shown in Fig. \ref{fig:BSMPdiv}(b).

\subsubsection{Matching Pennies}

Finally, we consider the Matching Pennies game with the more general payoff matrix shown in Table \ref{table:BSMPpayoff}.  In the case of the Matching Pennies we have that $a_{1} = e^{\lambda_{1}}, b_{1}= e^{-\lambda_{1}(1+\beta)}, a_{2} = e^{-\lambda_{2}}, b_{2} = e^{2\lambda_{2}}$ and equations \eqref{eq:qre} take the form:
\begin{equation}
\begin{aligned}
p^{1}_{c}(1+e^{\lambda_{1}(1-(1 + \beta) p^{2}_{c})}) &= 1\\
p^{2}_{c}(1+e^{\lambda_{2}(-1+2 p^{1}_{c})}) &= 1
\end{aligned}
\label{eq:mp}
\end{equation}
Equations \eqref{eq:mp} are again not symmetric when exchanging players and so the game is played twice on each edge, as also done for the Battle of the Sexes.  We can approximate the probabilities for strategy choice C to zero order by $p^{1}_{c} = p^{2}_{c} = 0.5$. If the rationalities are small then we can make the following first order approximations:
\begin{equation}
\begin{aligned}
p^{1}_{c} &\simeq \frac{1}{2}+\frac{\lambda_{1}(\beta-1)}{8}\\
p^{2}_{c} &\simeq \frac{1}{2}-\frac{\lambda_{1}\lambda_{2}(\beta-1)}{16}
\label{eq:mp1st}
\end{aligned}
\end{equation}

Table \ref{table:BSMPdiv} gives figures for the average divergence for different topologies and rationality functions, while Fig. \ref{fig:BSMPdiv}(a) shows the results obtained numerically from an optimisation procedure to minimise $-\rho$. The results are largely analogous to the Battle of Sexes, with one exception: The core-periphery graph that minimises the divergence has a core consisting of a quasi-regular graph, with $N_{c} = 168$ nodes with average degree $d_{h} \simeq 19$, and the periphery nodes form $(N-N_{c})/2 = 416$ pairs ($K_{2}$ graphs) completely disconnected from the core. Equation \eqref{eq:div_asym} can be applied with $\alpha = N_{c}d_{h}/(2M) \simeq 80\%$, and the results match with the entries in Table \ref{table:BSMPdiv}. If we allow isolated nodes to form during the optimisation, then the optimal network structure that minimises the average divergence is a quasi-regular graph with $N_{c} = 168$ and average degree $d_{h} \simeq 24$. Hence, the core in both types of optimisation for the Matching Pennies game has the same number of nodes; it is only the number of edges within the core that varies due to the connectivity constraint.

\section{Conclusion}
\label{sec:cc}

Previous work has proposed that scale-free networks maximise the overall "rationality" of a networked system of players \cite{Kasthurirathna2015}. Here we have shown that for the Prisoner's Dilemma game core-periphery graphs, composed of a core of connected hub nodes which each link to a single periphery node, have a significantly lower average Nash-LQRE divergence than scale-free networks. This result is supported by numerical experiments and also by an analytic approximation for the divergence of core-periphery graphs that reproduces numerical results well for all the different rationality functions considered. The accuracy of the analytic results can decrease due to the fact that in finite core-periphery graphs the hubs will always have a finite degree, and hence the approximations suffer. If no connectivity constraint is enforced during optimisation, optimal networks are found to consist of a core made up by a complete graph with all other nodes being isolated. 

For the Stag Hunt game, an analogous analytic treatment of the connectivity constrained case yields qualitatively similar results to the Prisoner's Dilemma, while the connectivity unconstrained optimisation leads to a regular graph.

Further, in contrast to \cite{Kasthurirathna2015}, we have demonstrated that highly heterogeneous degree distributions do not necessarily maximise system rationality for all classes of games. Analytic and simulation results show that, for the Battle of the Sexes and the Matching Pennies games, core-periphery graphs with cores that have a quasi-regular topology minimise the average Nash-LQRE divergence. If the connectivity of the network is not constrained in the optimisation procedure, then in the case of the Battle of the Sexes a graph with a strongly bi-modal degree distribution emerges, while for the Matching Pennies game we obtain a quasi-regular graph.

We can interpret the results for the Prisoner’s Dilemma or Stag Hunt games in a broader way. Cooperation is naturally associated with altruistic behaviour, whereas defection can be seen as a selfish course of action. Maximising system rationality, or minimising the difference between the probability distributions and the Nash equilibria, can thus be interpreted as promoting selfish behaviour within the system. The topology that emerges is one where a core consisting of roughly $1\%$ of the nodes in the network are part of over $90\%$ of the connections available, whereas each periphery node is as poorly connected as possible. If higher numbers of a node in a social network translates into greater influence/wealth, then we see that by maximising selfishness in a social network most of the power ends up concentrated in a very small set of players. This result echoes present concerns regarding wealth inequality, and furthermore, a similar core-periphery structure has been found to underlie the network of global corporate control, wherein a ``large portion of control flows to a small tightly-knit core of financial institutions'' \cite{Vitali2011}. Our results elaborate a possible mechanism for the emergence of such highly concentrated core-periphery structures in social networks.

\section*{Acknowledgements}
This work was supported by an EPSRC Doctoral Training Centre Grant (EP/G03690X/1). No new data was collected by this research.

\newpage
\bibliographystyle{unsrtnat}
\bibliography{bib}

\begin{thebibliography}{28}
\providecommand{\natexlab}[1]{#1}
\providecommand{\url}[1]{\texttt{#1}}
\expandafter\ifx\csname urlstyle\endcsname\relax
  \providecommand{\doi}[1]{doi: #1}\else
  \providecommand{\doi}{doi: \begingroup \urlstyle{rm}\Url}\fi

\bibitem[Cardell et~al.(1997)Cardell, Hitt, and Hogan]{Cardell1997}
J.B. Cardell, C.C. Hitt, and W.W. Hogan.
\newblock Market power and strategic interaction in electricity networks.
\newblock \emph{Resource and energy economics}, 19:\penalty0 109--137, 1997.

\bibitem[Nowak and Sigmund(2004)]{Nowak2004}
M.A. Nowak and K.~Sigmund.
\newblock Evolutionary dynamics of biological games.
\newblock \emph{Science}, 303:\penalty0 793--799, 2004.

\bibitem[De~Mesquita(2006)]{DeMesquita2006}
B.B. De~Mesquita.
\newblock Game theory, political economy, and the evolving study of war and
  peace.
\newblock \emph{American Political Science Review}, 100:\penalty0 637--642,
  2006.

\bibitem[Nash(1950)]{Nash1950}
J.~F. Nash.
\newblock Equilibrium points in n-person games.
\newblock \emph{Proc. Natl. Acad. Sci.}, 36:\penalty0 48--49, 1950.

\bibitem[Nell and Errouaki(2013)]{Nell2013}
E.~J. Nell and K.~Errouaki.
\newblock \emph{Rational econometric man: transforming structural
  econometrics}.
\newblock Edward Elgar, Cheltenham, UK, 2013.

\bibitem[McKelvey and Palfrey(1995)]{McKelvey1995}
R.~D. McKelvey and T.~R. Palfrey.
\newblock Quantal response equilibria for normal form games.
\newblock \emph{Game. Econ. Behav.}, 10:\penalty0 6--38, 1995.

\bibitem[Luce(2005)]{Luce2005}
R.D. Luce.
\newblock \emph{Individual choice behavior: A theoretical analysis}.
\newblock Dover, Mineola, NY, 2005.

\bibitem[Ochs(1995)]{Ochs1995}
J.~Ochs.
\newblock Games with unique, mixed strategy equilibria: An experimental study.
\newblock \emph{Games and Economic Behavior}, 10:\penalty0 202--217, 1995.

\bibitem[Goeree et~al.(2002)Goeree, Holt, and Palfrey]{Goeree2002}
J.~K. Goeree, C.~A. Holt, and T.~R. Palfrey.
\newblock Quantal response equilibrium and overbidding in private-value
  auctions.
\newblock \emph{Journal of Economic Theory}, 104:\penalty0 247--272, 2002.

\bibitem[Enriqueta and Thomas(2004)]{Enriqueta2004}
A.~Enriqueta and R.~P. Thomas.
\newblock The effect of candidate quality on electoral equilibrium: An
  experimental study.
\newblock \emph{The American Political Science Review}, 98:\penalty0 77--90,
  2004.

\bibitem[McKelvey and Patty(2006)]{McKelvey2006}
R.~D. McKelvey and J.~W. Patty.
\newblock A theory of voting in large elections.
\newblock \emph{Games and Economic Behavior}, 57:\penalty0 155--180, 2006.

\bibitem[Rogers et~al.(2009)Rogers, Palfrey, and Camerer]{Rogers2009}
B.~W. Rogers, T.~R. Palfrey, and C.~F. Camerer.
\newblock Heterogeneous quantal response equilibrium and cognitive hierarchies.
\newblock \emph{Journal of Economic Theory}, 144:\penalty0 1440--1467, 2009.

\bibitem[Choi et~al.(2012)Choi, Gale, and Kariv]{Choi2012}
S.~Choi, D.~Gale, and S.~Kariv.
\newblock Social learning in networks: a quantal response equilibrium analysis
  of experimental data.
\newblock \emph{Review of Economic Design}, 16:\penalty0 135--157, 2012.

\bibitem[Nowak and May(1992)]{Nowak1992}
M.~A. Nowak and R.~M. May.
\newblock Evolutionary games and spatial chaos.
\newblock \emph{Nature}, 359:\penalty0 826--829, 1992.

\bibitem[Szab\'{o} and Borsos(2016)]{Szabo2016}
G.~Szab\'{o} and I.~Borsos.
\newblock Evolutionary potential games on lattices.
\newblock \emph{Physics Reports}, 624:\penalty0 1--60, 2016.

\bibitem[Szab\'{o} and Fath(2007)]{Szabo2007}
G.~Szab\'{o} and G.~Fath.
\newblock Evolutionary games on graphs.
\newblock \emph{Physics reports}, 446:\penalty0 97--216, 2007.

\bibitem[Perc et~al.(2013)Perc, G\'{o}mez-Garde\~{n}es, Szolnoki, Floría, and
  Moreno]{Perc2013}
M.~Perc, J.~G\'{o}mez-Garde\~{n}es, A.~Szolnoki, L.M. Floría, and Y.~Moreno.
\newblock Evolutionary dynamics of group interactions on structured
  populations: a review.
\newblock \emph{Journal of The Royal Society Interface}, 10, 2013.

\bibitem[Santos and Pacheco(2005)]{Santos2005}
F.~C. Santos and J.~M. Pacheco.
\newblock Scale-free networks provide a unifying framework for the emergence of
  cooperation.
\newblock \emph{Phys. Rev. Lett.}, 95:\penalty0 098104, 2005.

\bibitem[Zimmermann et~al.(2004)Zimmermann, Eguíluz, and
  San~Miguel]{Zimmermann2004}
M.G. Zimmermann, V.M. Eguíluz, and M.~San~Miguel.
\newblock Coevolution of dynamical states and interactions in dynamic networks.
\newblock \emph{Physical Review E}, 69:\penalty0 065102, 2004.

\bibitem[Szolnoki et~al.(2008)Szolnoki, Perc, and Danku]{Szolnoki2008}
A.~Szolnoki, M.~Perc, and Z.~Danku.
\newblock Making new connections towards cooperation in the prisoner's dilemma
  game.
\newblock \emph{EPL}, 84:\penalty0 50007, 2008.

\bibitem[Kasthurirathna and Piraveenan(2015)]{Kasthurirathna2015}
D.~Kasthurirathna and M.~Piraveenan.
\newblock Emergence of scale-free characteristics in socio-ecological systems
  with bounded rationality.
\newblock \emph{Nature Scientific reports}, 5:10448, 2015.

\bibitem[Kasthurirathna et~al.(2016{\natexlab{a}})Kasthurirathna, Piraveenan,
  and Uddin]{Kasthurirathna2016a}
D.~Kasthurirathna, M.~Piraveenan, and S.~Uddin.
\newblock Modeling networked systems using the topologically distributed
  bounded rationality framework.
\newblock \emph{Complexity}, 21:\penalty0 123--137, 2016{\natexlab{a}}.

\bibitem[Kasthurirathna et~al.(2016{\natexlab{b}})Kasthurirathna, Harr\'{e},
  and Piraveenan]{Kasthurirathna2016b}
D.~Kasthurirathna, M.~Harr\'{e}, and M.~Piraveenan.
\newblock Optimising influence in social networks using bounded rationality
  models.
\newblock \emph{Social Network Analysis and Mining}, 6:\penalty0 54,
  2016{\natexlab{b}}.

\bibitem[Li et~al.(2005)Li, Alderson, Doyle, and Willinger]{Li2005}
L.~Li, D.~Alderson, J.~C. Doyle, and W.~Willinger.
\newblock Towards a theory of scale-free graphs: Definition, properties, and
  implications.
\newblock \emph{Internet Mathematics}, 2:\penalty0 431--523, 2005.

\bibitem[Brede(2010{\natexlab{a}})]{Brede2010a}
M.~Brede.
\newblock Coordinated and uncoordinated optimization of networks.
\newblock \emph{Physical Review E}, 81:\penalty0 066104, 2010{\natexlab{a}}.

\bibitem[Brede(2010{\natexlab{b}})]{Brede2010b}
M~Brede.
\newblock Optimal synchronization in space.
\newblock \emph{Physical Review E}, 81:\penalty0 025202, 2010{\natexlab{b}}.

\bibitem[Barab\'{a}si and Albert(1999)]{Barabasi1999}
A.L. Barab\'{a}si and R.~Albert.
\newblock Emergence of scaling in random networks.
\newblock \emph{Science}, 286:\penalty0 509--512, 1999.

\bibitem[Vitali et~al.(2011)Vitali, Glattfelder, and Battiston]{Vitali2011}
S.~Vitali, J.~B. Glattfelder, and S.~Battiston.
\newblock The network of global corporate control.
\newblock \emph{PloS one}, 6:\penalty0 e25995, 2011.

\end{thebibliography}

\end{document}